# Impacts of Phase Noise on M-ary QAM THz Wireless Communications

Bowen Liu*
*Department of Electronics and Electrical Engineering*
*Keio University*
Yokohama, Japan
b.liu@phot.elec.keio.ac.jp

Takasumi Tanabe
*Department of Electronics and Electrical Engineering*
*Keio University*
Yokohama, Japan
takasumi@elec.keio.ac.jp

*Abstract*—THz technology is positioned as a key enabler for next-generation wireless links due to the extensive untapped bandwidth and inherent compatibility with silicon photonics. Here, the phase noise of THz sources is modeled to characterized its impacts on a QAM-based wireless communication system. Particularly, common and instantaneous phase errors are unveiled and presented via constellation diagrams. In addition, tolerance of phase noise under different M-ary modulation formats is explored. Meanwhile, a linear correlation between instantaneous component and EVM is revealed to define the error-free boundary, where microcomb-driven scheme is highlighted as a highly promising candidate for error-resilient THz links under advanced M-QAM formats. The results intuitively illustrate the impacts of phase noise and offer practical insights for refining technical details on physical-layer protocols.

## I. Introduction

Digital economy has initiated a new wave of data-driven industrial revolution. In 2022, the mobile data users alone in the Asia-Pacific region surged by 3.1 billion, with monthly average data consumption increasing by 11GB [1]. However, the shortage in bandwidth resources severely restrict the capacity of mobile data traffic, and is becoming an increasingly urgent social challenge. The average wireless connection speed remained at 28.5 Mbps in the Asia-Pacific region, where approximately 47% of the accesses experience a peak speed below 100 Mbps [1]. The accelerating growth of mobile data traffic has triggered an urgency for ultrafast, robust, widely affordable and accessible wireless communication technologies, which are essential for societal and individual development.

Terahertz (THz) waves, as defined at frequencies ranging from 100 GHz to 10 THz, are considered a highly prospective playground for future ultra-high speed links [2,3]. Particularly, around 44-GHz bandwidth resources are available for wireless communications in the 300-GHz band [4], which is 110 times the currently defined 400 MHz in 3GPP 5G NR protocols [5]. Owing to the abundant unexploited bandwidth and inherent compatibility with photonic integrated devices, THz wireless links have been intensively attracted to address the rapidly growing demand for ultrafast data rates, large capacity, and high energy efficiency [6]. In 2011, 100 Gbit/s 16 quadrature amplitude modulation (QAM) wireless transmission was demonstrated in 75-110 GHz band [7]. Higher data rates was further achieved to 160 Gbit/s within 300-500 GHz band through eight 25-GHz separated channels [8]. T. Nagatsuma's group reported a 220 Gbit/s 32QAM transmission over a 214-m free-space link [9]. However, early demonstrations relied on conventional laser sources or frequency-multiplied electronic resonators, whose phase noise characteristics impose fundamental limitations on the feasibility of advanced modulation formats beyond 32QAM. Recent efforts have explored alternative approaches such as quantum-dash lasers with external coherent feedback, achieving 12 Tbit/s 32QAM THz links [10]. In addition, self-injection-locked microring resonators have been employed to enable ultra-high-speed THz wireless communication using 32QAM modulation [11]. Although the ultra-high speed potential in the THz band has been widely demonstrated, and ultra-low phase noise sources such as silicon photonic microresonators are anticipated to support advanced modulation formats comparable to existing 5G (256QAM) and Wi-Fi 6 (1024QAM) technologies, quantitative understanding of phase noise impact remain limited. This gap poses challenges for the design and scalability of future high-capacity THz-wireless systems.

In this paper, we present a modeling framework for phase noise in THz sources and perform an intuitive analysis of its impact on signal quality. Phase noise degrades the effective signal-to-noise ratio (SNR), manifesting as steep deterioration in bit error rate (BER) and error vector magnitude (EVM). More importantly, its high-frequency jitters severely constrain the achievable performance celling and remain uncorrectable via standard compensation techniques. The phase noise tolerance across different QAM formats has been quantitatively revealed, offering valuable insights for the development of physical-layer protocols in future THz wireless communication systems.

## II. Principle

### A. System-Level Simulation Model.

The simulation is based on a typical physical-layer architecture of a single-carrier THz wireless communication system, as illustrated in Fig. 1. It comprises a transmitter, a receiver, and an additive white Gaussian noise (AWGN) channel. The baseband data are first encoded and digitally modulated using M-ary QAM. These symbols are then

This work was supported by JST, CRONOS, Japan (JPMJCS24N7).

converted into analog waves and up-converted onto a THz carrier through photonic mixing with a local oscillator (LO) signal, which introduces phase noise. The THz signal is transmitted through a lensed horn antenna and propagates over a simplified flat-fading channel, where frequency-selective fading is neglected. At the receiver, the signal is down-converted, digitized, and processed via digital signal processing (DSP). Finally, the system performance is evaluated in terms of BER and EVM.

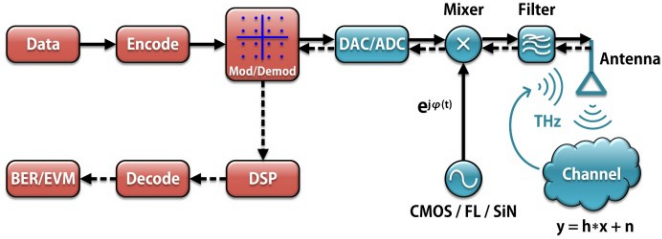

Fig. 1. Processing chain of the system-level simulation model.

*B. Modeling of Location Oscillator's Phase Noise*

Phase noise of LOs is modeled based on their single-sideband (SSB) power spectral density (PSD) and normalized to 300 GHz band. As an example, Fig. 2 illustrates the phase noise characteristics of three typical THz sources—a complementary metal oxide semiconductor (CMOS) oscillator defined by 3GPP 5G NR protocols [5], a stimulated Brillouin scattering (SBS) fiber laser (FL) [12], and a silicon nitride (SiN) microcomb [13]. The original PSD are resampled according to the simulation's sampling rate and multiplied by a random phase spectrum $\theta(f)$, uniformly distributed over $-\pi$ to $+\pi$, to construct the complete complex phase noise spectrum $O(f)$ [14].

$$O(f) = \begin{cases} \sqrt{\dfrac{L(f)}{2}} e^{j\theta(f)}, & f > 0 \\ \sqrt{\dfrac{L(-f)}{2}} e^{-j\theta(f)}, & f < 0 \\ 1, & f = 0 \end{cases} \quad (1)$$

An inverse Fourier transform is then applied to generate the corresponding temporal phase noise $\varphi(t)$, which can be directly imposed onto the modulated signal without convolution operator in the frequency domain.

$$\varphi(t) = Re\{IFT[O(f)]\} \quad (2)$$

Figure 2(a) shows the SSB PSD of the three sources, with corresponding standard deviations (STD, **σ**) of 0.2577, 0.0108, and 0.0066, respectively. Figure 2(b) presents the synthesized time-domain phase noise waveform of the SiN microcomb, while Fig. 2(c) shows its probability density function (PDF), confirming a Gaussian distribution centered at zero with **σ** = 0.0066. To comprehensively assess phase noise tolerance, additional simulations are conducted over a range of STDs.

## III. RESULTS AND DISCUSSION

*A. Phase Noise Tolerance of probability estimated BER for M-ary QAM Formats*

We first investigate the relationship between phase noise and BER in the simulated THz communication system. Phase noise introduces random phase fluctuations and instantaneous frequency jitter of the LO, which leads to irregular symbol timing delays, inter-symbol power leakage, and degradation in effective SNR. These effects cause severe signal distortion and increase the BER, ultimately lowering the system performance ceiling. As illustrated in Fig. 3(a), we evaluate 256QAM BER under various phase noise STDs. The BER surface reveals a steep rise with increasing σ and decreasing $E_b/N_0$, forming a topography with high BER in the northwest and low BER in the southeast. The three representative LOs introduced in Fig. 2 are mapped onto this surface. It should be noted that electronic oscillators, including CMOS and voltage-controlled oscillator (VCO), always require frequency multiplication to reach the THz band (*e.g.* 300 GHz). This results in an effectively more severe phase noise level in correspondence with frequency multiple in practice.

Figure 3(b) shows the BER dependence on $E_b/N_0$ for various phase noise levels. As $E_b/N_0$ increases, BER decreases with different convergence rate: a lower-level of phase noise yields faster BER improvement, reflecting stronger system robustness. This implies that phase noise reduces the usable SNR in actual. To achieve a BER comparable to the ideal specification requires significantly higher SNR, which imposes tighter constraints on power budget and device noise figure of system design. Additionally, Fig. 3(c) provides the BER floor versus phase noise under specific $E_b/N_0$ (below 28 dB), where a trend inflection is observed, indicating a phase noise tolerance. Below this threshold, further reduction in phase noise yields greatly diminishing returns in BER improvement.

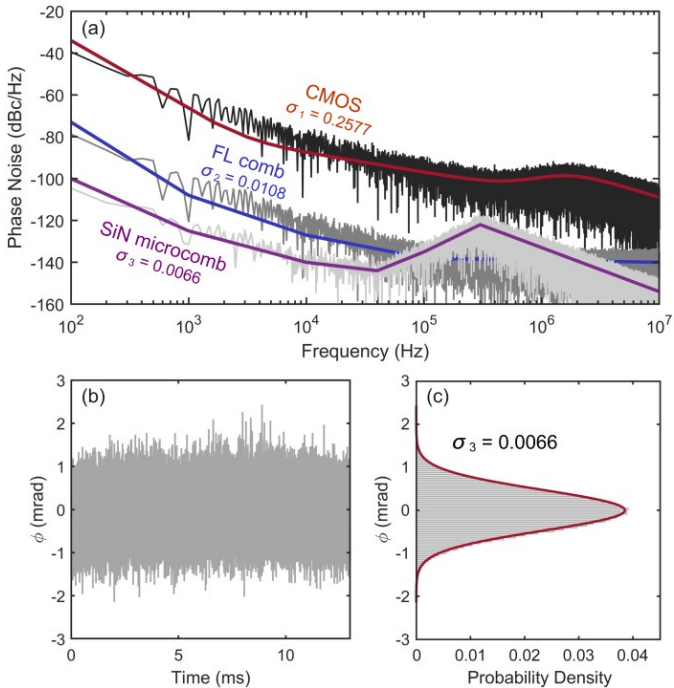

Fig. 2. (a) SSB phase noise spectra of three THz sources: CMOS oscillator, FL comb, and SiN microcomb, with fitted standard deviations $\sigma_1$, $\sigma_2$, and $\sigma_3$; (b) Time-domain phase noise for SiN microcomb; (c) Corresponding PDF.

## B. Impacts of Common and Instantaneous Phase Noise

Whereas a practical systems cannot indefinitely increase SNR to compensate for high phase noise. Meanwhile, BER estimation becomes statistically unreliable in extremely low-error regimes. Therefore, we further analyze the phase noise influences from the perspective of constellation distortion and EVM. Constellation diagrams of the decoded signals using different THz sources are shown in Fig. 5(a) to 5(c), showing a root-mean-square (RMS) EVM of 30.75%, 3.30% and 3.16% without common phase recovery (CPR), respectively. In Fig. 5(a), the CMOS-driven system exhibits two distinct types of phase noise impacts: a global rotation of the entire constellation (attributed to low-frequency phase drift) and local spinning of individual symbols (due to high-frequency jitter). This decomposition becomes more evident in Fig. 5(d), where AWGN is removed to isolate the impact of phase noise. The low-frequency component introduce a slow and time-correlated common phase error (CPE) that accumulatively shift all temporal symbols collectively. It can be largely mitigated by physical phase-locked loops (PLLs) or digital CPR algorithms. In contrast, the high-frequency component result in an instantaneous and random symbol-level distortions that cannot be effectively tracked or compensated as shown in Fig. 5(e), where residual phase errors (~4.26%) persist even after CPR.

In comparison, the fiber laser and SiN demonstrate superior stability in both common and instantaneous phase noise. Their constellation points remain tightly clustered even without CPR, as shown in Fig. 5(b) and (c). To quantitatively evaluate the constellation fuzziness induced by instantaneous phase noise, the RMS EVM is computed and converted into its 3σ error radius $r_{3\sigma}$, which defines the circular boundary enclosing 99.7% of blurred symbols' distribution. As illustrated in Fig. 5(f), the dashed blue circles present the tolerance for symbols. Once circles begin to overlap, energy leakage between neighboring symbols occurs, marking the onset of decision errors.

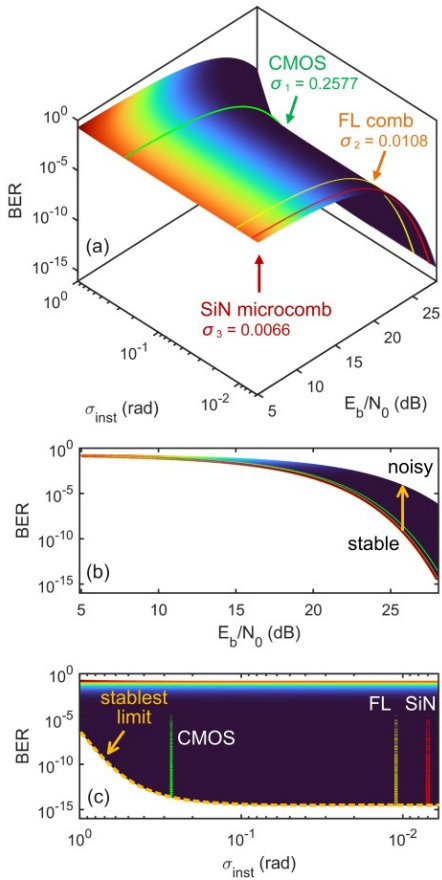

Fig. 3. 256QAM BER: (a) the relationships of BER, SNR and phase noise STD; (b) BER versus SNR; (c) BER versus phase noise STD.

Furthermore, the dependence of BER on phase noise for various M-ary QAM formats is summarized in Fig. 4 at 20-dB $E_b/N_0$. Higher-order modulation formats exhibit poorer BER performance and lower tolerance to phase noise under the same SNR, indicating tighter constraints on LO stability. Particularly, we highlight two commonly used BER thresholds: $1\times10^{-9}$ for no forward error correction (w/o FEC) and $2.2\times10^{-3}$ for ultra-FEC (UFEC) systems. For instance, the maximum tolerable phase noise for 16QAM and 32QAM to achieve a BER below $1\times10^{-9}$ are identified as points $a$ (σ ≤ 0.903) and $b$ (σ ≤ 0.565), respectively. Similarly, 128QAM and 256QAM can achieve a UFEC-limit BER of $2.2\times10^{-3}$ at points $c$ (σ ≤ 0.953) and $d$ (σ ≤ 0.577), respectively.

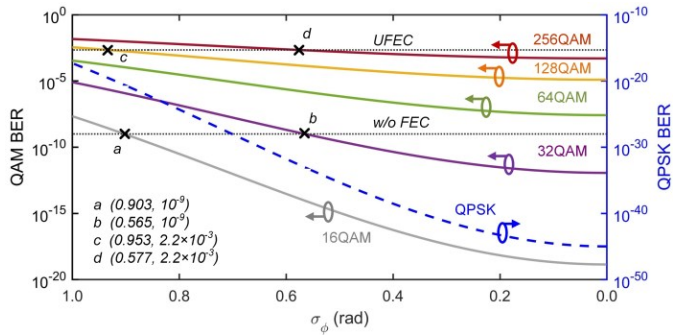

Fig. 4. BER versus phase noise STD for various M-ary QAM at 20-dB SNR.

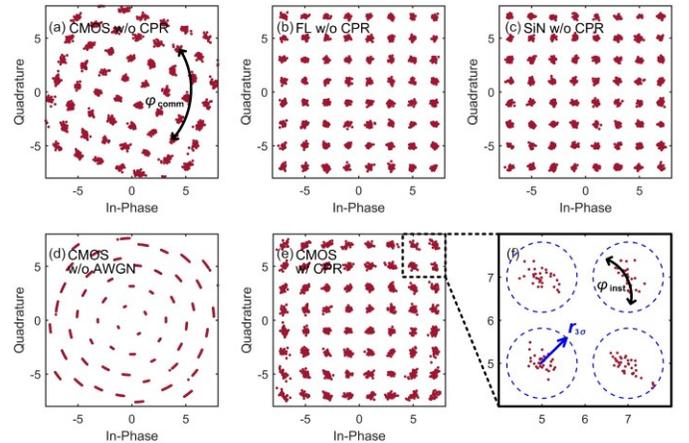

Fig. 5. 64QAM constellation diagram at 30-dB SNR without CPR using (a) CMOS, (b) fiber laser, and (c) SiN; (d) CMOS case with AWGN removed; (e) CMOS with CPR showing residual intra-symbol distortion; (f) Enlarged view of symbol clusters with 3σ error vector power radius.

## C. Tolerance of Unrecoverable Instantaneous Phase Noise for M-ary QAM Formats

We further investigate the impact of unrecoverable instantaneous phase noise that remains after CPR. The 64QAM

constellation diagrams in Fig. 6(a) to 6(c) represent the residual distortion over a complete subframe period for three examined LOs. Due to the increased symbol duration, the variance of instantaneous phase jitter also increases, resulting in more pronounced intra-symbol rotational errors, particularly in the CMOS-driven system. In contrast, the fiber laser and SiN microcomb exhibit tightly clustered symbols with clearly distinguishable constellation points, reflecting their superior phase noise stability. Here, the error vectors $e_k$ for each received symbol are calculated, and then a RMS EVM is derived.

$$e_k = (I_k - \bar{I}_k)^2 + (Q_k - \bar{Q}_k)^2 \qquad (3)$$

$$EVM_{RMS} = 100\% \times \sqrt{\frac{\frac{1}{N}\sum e_k}{\frac{1}{N}\sum(I_k^2 + Q_k^2)}} \qquad (4)$$

Where $I_k$ and $Q_k$ respectively stand for the real and imaginary part of the $k$-th complex symbol.

As shown in Fig. 6(d), RMS EVM exhibits a strong linear correlation with the standard deviation $\sigma_{inst}$ of instantaneous phase noise for 64QAM. Under normalized constellation scaled by the minimum inter-symbol distance of 2, the 3σ error boundary—defined as the radius within which 99.7% of symbols fall—is reached when RMS EVM is 5.12%. This corresponds to a threshold phase noise level of a $\sigma_{inst}$ equals 0.0042, beyond which significant symbol overlap and decision errors occur. In Fig. 6(d), the CMOS source lies well beyond the threshold, with an RMS EVM of approximately 10.71%; while the fiber laser and SiN remain within the tolerance region, yielding 3.27% and 3.17% RMS EVM respectively.

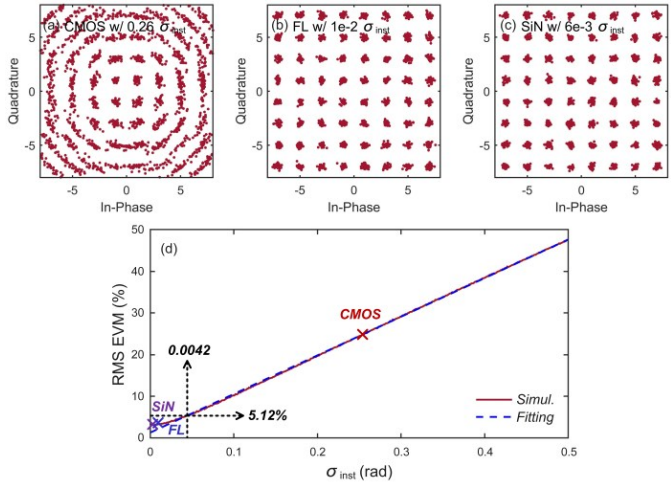

Fig. 6. Unrecoverable instantaneous phase error: 30-dB SNR 64QAM constellation diagram using (a) CMOS, (b) fiber laser, and (c) SiN microresonator. (d) Relationship between RMS EVM and $\sigma_{inst}$.

Furthermore, general conclusion is extended to analyze various M-ary QAM scenes via the 3σ instantaneous error vector radius $r_{3\sigma}$.

$$r_{3\sigma} = 3 \times EVM_{RMS}\sqrt{P_M} = EVM_{RMS}\sqrt{6(M-1)} \qquad (5)$$

As shown in Fig. 7, $r_{3\sigma}$ indicates the circular region enclosing 99.7% of constellation points due to instantaneous phase noise. When $r_{3\sigma} > 1$, adjacent symbol boundaries begin to overlap, introducing a risk of inter-symbol interference and symbol decision errors. The black markers (*a* to *d*) specify the tolerance thresholds for 128QAM, 64QAM, 32QAM, and 16QAM, respectively. Notably, 256QAM exceeds this threshold even at very low phase noise levels, implying that achieving error-free performance requires a higher SNR than 30 dB. Moreover, when $r_{3\sigma}$ exceeds $\sqrt{2}$, diagonal symbol regions begin to overlap, leading to more difficult-to-correct decision errors. These observations highlight the increasingly stringent phase noise requirements as modulation order increases and underscore the importance of oscillator quality in enabling high-order QAM for THz wireless links.

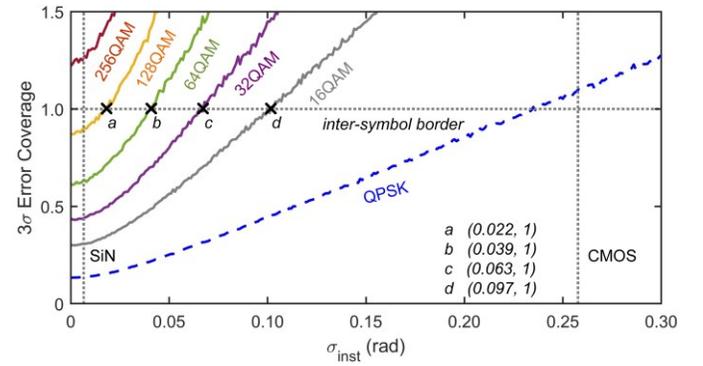

Fig. 7. 3σ instantaneous error vector radius $r_{3\sigma}$ versus phase noise $\sigma_{inst}$ for various M-ary QAM modulations at 30-dB SNR.

## IV. CONCLUSION

In this work, we have systematically investigated the tolerance of phase noise in THz wireless communication systems under M-ary QAM modulation. A unified modeling framework is developed to simulate the intrinsic phase noise of LOs, where typical THz sources including CMOS, fiber laser, and SiN microresonator are discussed for reference. More generally, different levels of phase noise are featured by STD, and their impacts on system performance are quantitatively assessed through BER and EVM analysis. We revealed that unrecoverable instantaneous phase noise imposes fundamental limitations on error performance, particularly for high-order modulation formats, and identified critical thresholds based on 3σ error vector radius. Linear correlations between RMS EVM and phase noise variances enable the derivation of practical tolerance boundaries, with microcombs demonstrating significant advantages in maintaining constellation integrity. In additional, increasing stringency of phase noise requirements at higher modulation orders is highlighted through exploring the symbol overlap determination across different M-ary QAM schemes. These results provide valuable insights for guiding the design of physical-layer protocols in future high-capacity THz wireless systems, emphasizing the importance of ultra-low phase noise sources such as silicon photonic microcombs.

ACKNOWLEDGMENT

This work was supported by JST, CRONOS, Japan Grant Number JPMJCS24N7.